\documentclass[aps,prb,reprint,superscriptaddress,showpacs,showkeys]{revtex4-1}

\usepackage{graphicx}
\usepackage{dcolumn}
\usepackage{bm}
\usepackage{color}

\begin{document}


\title[Sample title]{Enhancement of the spin Hall voltage in a reverse-biased planar pn-junction}

\author{L.~N\'{a}dvorn\'{i}k}
 \email{nadvl@fzu.cz}
 \affiliation{Institute of Physics ASCR, v.v.i., Cukrovarnick\'{a} 10, 16253 Praha 6, Czech Republic}
 \affiliation{Faculty of Mathematics and Physics, Charles University, Ke Karlovu 3, 12116 Praha 2, Czech Republic}
\author{K.~Olejn\'{i}k}
 \affiliation{Institute of Physics ASCR, v.v.i., Cukrovarnick\'{a} 10, 16253 Praha 6, Czech Republic}
\author{P.~N\v{e}mec}
 \affiliation{Faculty of Mathematics and Physics, Charles University, Ke Karlovu 3, 12116 Praha 2, Czech Republic}
\author{V.~Nov\'{a}k}
 \affiliation{Institute of Physics ASCR, v.v.i., Cukrovarnick\'{a} 10, 16253 Praha 6, Czech Republic}
\author{T.~Janda}
 \affiliation{Faculty of Mathematics and Physics, Charles University, Ke Karlovu 3, 12116 Praha 2, Czech Republic}
 \affiliation{Institute of Physics ASCR, v.v.i., Cukrovarnick\'{a} 10, 16253 Praha 6, Czech Republic}
 \author{J.~Wunderlich}
 \affiliation{Institute of Physics ASCR, v.v.i., Cukrovarnick\'{a} 10, 16253 Praha 6, Czech Republic}
 \affiliation{Hitachi Cambridge Laboratory, J. J. Thomson Avenue, CB3 0HE Cambridge, UK}
 \author{F.~Troj\'{a}nek}
 \affiliation{Faculty of Mathematics and Physics, Charles University, Ke Karlovu 3, 12116 Praha 2, Czech Republic}
\author{T.~Jungwirth}
 \affiliation{Institute of Physics ASCR, v.v.i., Cukrovarnick\'{a} 10, 16253 Praha 6, Czech Republic}
 \affiliation{School of Physics and Astronomy, University of Nottingham, Nottingham NG7 2RD, UK}
 


\date{\today}

\begin{abstract}
{We report an experimental demonstration of a local amplification of the spin Hall voltage using an expanding depletion zone at a pn-junction in GaAs/AlGaAs Hall-bar microdevices. It is demonstrated that the depletion zone can be spatially expanded by applying reverse bias by at least 10~$\mu$m at low temperature. In the depleted regime, the spin Hall signals reached more than one order of magnitude higher values than in the normal regime at the same electrical current flowing through the micro-device. It is shown that the pn-bias has two distinct effects on the detected spin Hall signal. It controls the local drift field at the Hall cross which is highly non-linear in the pn-bias due to the shift of the depletion front. Simultaneously, it produces a change in the spin-transport parameters due to the non-linear change in the carrier density at the Hall cross with the pn-bias.}

\end{abstract}

\pacs{72.25.Dc, 72.25.Fe}
\keywords{Spin Hall effect, pn-junction, depletion, optical orientation in semiconductors, polarimeter}
\maketitle



\section{Introduction}
During the last decade, the direct and inverse spin Hall effects (SHE) \cite{kato2004,wunderlich2004,wunderlich2005, valenzuela2006, saitoh2006, zhao2006} have been established as important tools in a wide variety of spintronic structures, where they act as generators or detectors of spin polarized currents in semiconductors and metals \cite{kato2004,wunderlich2004,wunderlich2005, valenzuela2006, saitoh2006, zhao2006, jungwirth2012}, in systems with a ferromagnetic layer as a trigger of magnetization reversal \cite{miron2011,liu2012}, or as electric polarimeters sensitive to the helicity of incoming light \cite{wunderlich2009,nadvornik2015}. Recently, a significant attention has been  also focused on concepts of logical spintronic devices based on the inverse spin Hall effect (ISHE) \cite{jungwirth2012}. In these proposals, the output ISHE signals are altered either using variations in the longitudinal drift biases \cite{nadvornik2015,olejnik2012}, or by using electric gates \cite{wunderlich2010}.

In this paper, we make use of both approaches by employing a depleted zone created by a lateral pn-junction in a two-dimensional electron gas (2DEG). We benefit from the fact that in the low-dimensional structures, unlike the bulk systems, the electrostatic depletion can be highly expanded in space by several microns by applying bias across a pn-junction \cite{gurugubelli2015}. 
In such a two-terminal transistor\cite{wang2014}, the relationship between the pn-bias and the local drift field at a Hall cross can be highly non-linear due to the propagation of the depletion front.
The presented results show that this unique transistor-like effect amplifies the ISHE voltages, detected at series of Hall crosses, by more than a factor of 30  with respect to a normal drift bias without the depletion effect.
It is shown that the position of the depletion front can be controlled with a sub-$\mu$m resolution by the pn-bias and that the ISHE sensor can be correspondingly switched between the high and low sensitivity operation.
Consistently with the previously reported observation \cite{matsuzaka2009}, we will finally discuss  the spin-related parameters with respect to the carrier depletion to due to the pn-junction gating.  
%

\section{Setup and sample}
\begin{figure}
\includegraphics[width=1.0\columnwidth]{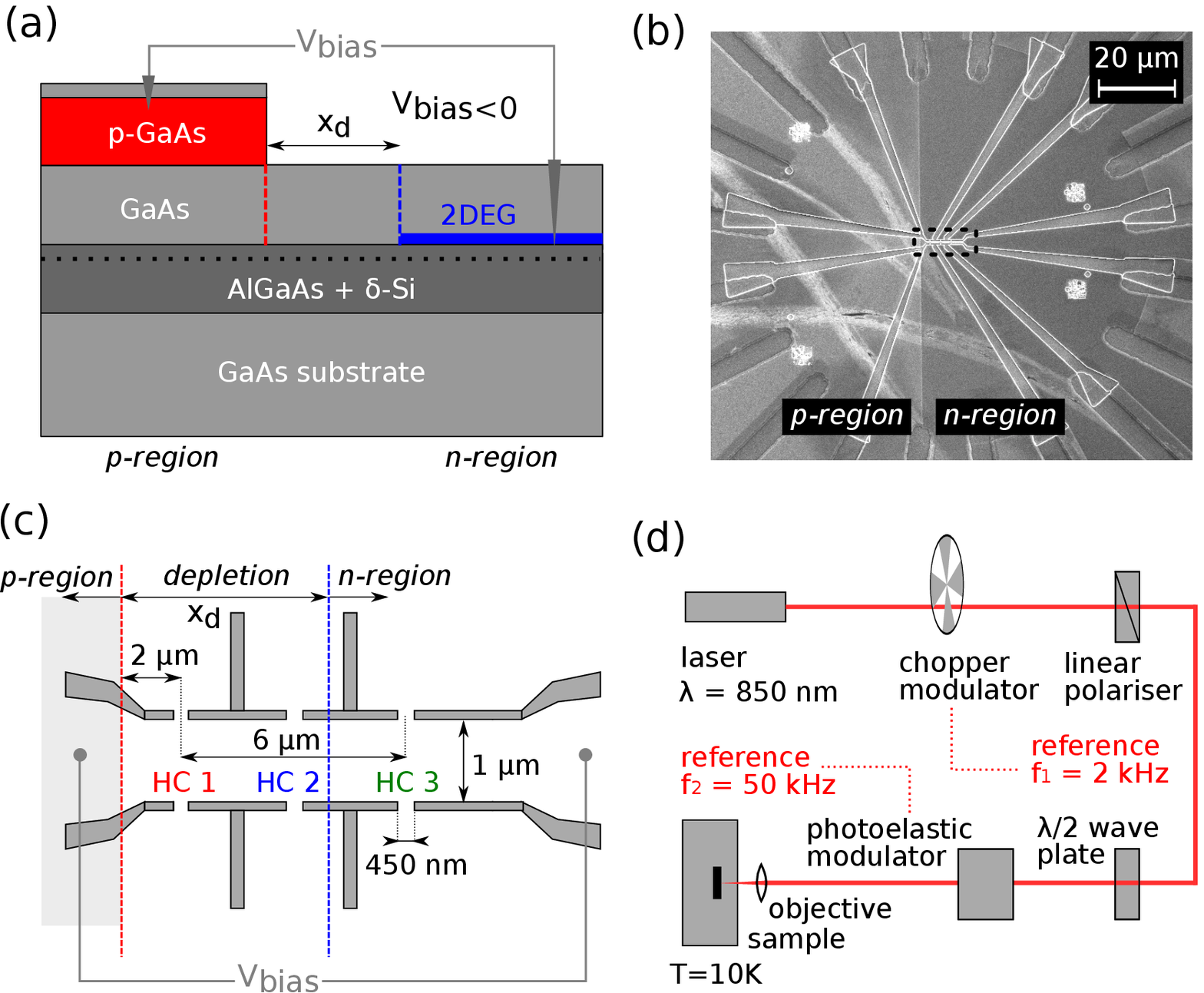}
\caption{\label{fig1} (Color online) (a) The layer composition of the sample, showing the created 2DEG (blue layer) and the p-doped layer (red layer). The $\delta$-Si doping layer and the formed pn-junction at $V_{\mathrm{bias}}=0$ are depicted by the black dotted and red dashed line, respectively, and $x_d$ is the width of the depleted zone for a significant reverse bias $V_{\mathrm{bias}}<0$, denoted by the blue dashed line. (b)~A~micro-image from a scanning electron microscope of the device showing the p- and n-region as the areas with different shades of grey. (c)~The sketch of the device design with depicted Hall crosses and dimensions (the region highlighted by the black dashed line in (b)). The red and blue dashed lines represent the depletion regions correspondingly to (a). The sketch is not to scale. (d)~The experimental setup with its two modulators: the chopper modulator and the PEM, operating at reference frequencies $f_1$ and $f_2$, respectively.}
\end{figure}

The experiments were performed at 10~K on a AlGaAs/GaAs-based heterostructure containing a 2DEG and a quasi-lateral pn-junction, as sketched in Fig.~\ref{fig1}(a). A sequence of silicon-doped Al$_{0.3}$Ga$_{0.7}$As layer (580~nm, $n_{Si}=9\times10^{11}$~cm$^{-2}$) and undoped GaAs layer (90~nm) was deposited by the molecular beam epitaxy on top of a semi-insulating GaAs substrate. The 2DEG, formed between these layers, had the moderate low-temperature mobility $\mu\approx1.6\times10^4$~cm$^2$V$^{-1}$s$^{-1}$ and the sheet electron density $n\approx8\times10^{11}$~cm$^{-2}$ after the light illumination. On top of it another carbon-doped GaAs layer (50~nm, $n_C=2\times10^{18}$~cm$^{-3}$) was deposited to create the p-region, followed by an undoped GaAs capping layer (10~nm). The last two layers were wet-etched out from a part of sample surface in order to create an n-region containing the unperturbed 2DEG with a quasi-lateral pn-junction at the etching edge (the red dashed curve in Fig.~\ref{fig1}(a) and the contrast line in the scanning electron microscope image in Fig.~\ref{fig1}(b)). Both the p- and n-regions were contacted using the Au/Cr and AuGeNi metalization, respectively, which allowed us to apply a bias voltage over the pn-junction. The corresponding I/V characteristic of the pn-junction is plotted in Fig.~\ref{fig2}(d), where the forward (positive) and reverse (negative) diode-like biasing regimes are clearly distinguishable. Here, $V_{\mathrm{bias}}$ and $I_{dc}$ are the voltage bias between the p- and n-region and the dc current flowing through the system, respectively. 

In order to accomplish the anticipated ISHE measurements, the heterostructure was surface-patterned as depicted in Figs.~\ref{fig1}(b,c). The Hall-bar design of marked dimensions was dry-etched in such a way that its left opening was positioned over the pn-junction (the red dashed line). The three Hall crosses (HCs) along the Hall-bar, HC~1, HC~2 and HC~3, were located at distances of 2, 5 and 8~$\mu$m away from the edge of the pn-junction and allowed us to observe the expansion of the depleted zone through the bar when the system is reverse-biased.

The experimental setup is shown in Fig.~\ref{fig1}(d). It combines the optical spin-injection via the optical orientation and the lock-in electrical detection of the ISHE voltage \cite{agranovich1984,nadvornik2015,wunderlich2009}. A continuous-wave Ti:sapphire laser is used to generate laser light of wavelength 850~nm, the polarization of which is, after setting its intensity to 100~$\mu$W, changed to circular by a set of a linear polarizer, a $\lambda/2$ wave-plate and a photo-elastic modulator (PEM) operated in the $\lambda/4$ mode. The light is then collected by a high-quality infra-red objective with 20$\times$ magnification and focused by it to the sample surface, where it forms a Gaussian spot with full width at half maximum (FWHM) $\sim2$~$\mu$m (estimated by the scanning knife-edge technique\cite{nemoto1989}). The objective is placed on a 3D piezo-electric stage that facilities scanning of the laser spot over the device with a sub-$\mu$m precision. The real time spot position with respect to the device was monitored by a CCD camera on a laser beam back-reflected from the sample. The light was double-modulated by the intensity modulator (the chopper wheel) and PEM (switching of the circular polarization between $\sigma_+$ and $\sigma_-$), operating at frequencies $f_1=2$~kHz and $f_2=50$~kHz, respectively. The double modulation technique enabled us to measure simultaneously the photo-current $I_{pc}$ at frequency $f_1$, which refers to the light-induced variations of the dc current $I_{dc}$ due to the generation of extra photo-carriers, and the ISHE voltages at $f_2$, which are dependent on the helicity of the circular polarization of the incoming light.

\section{Results}
\begin{figure}[htb]
\includegraphics[width=1.0\columnwidth]{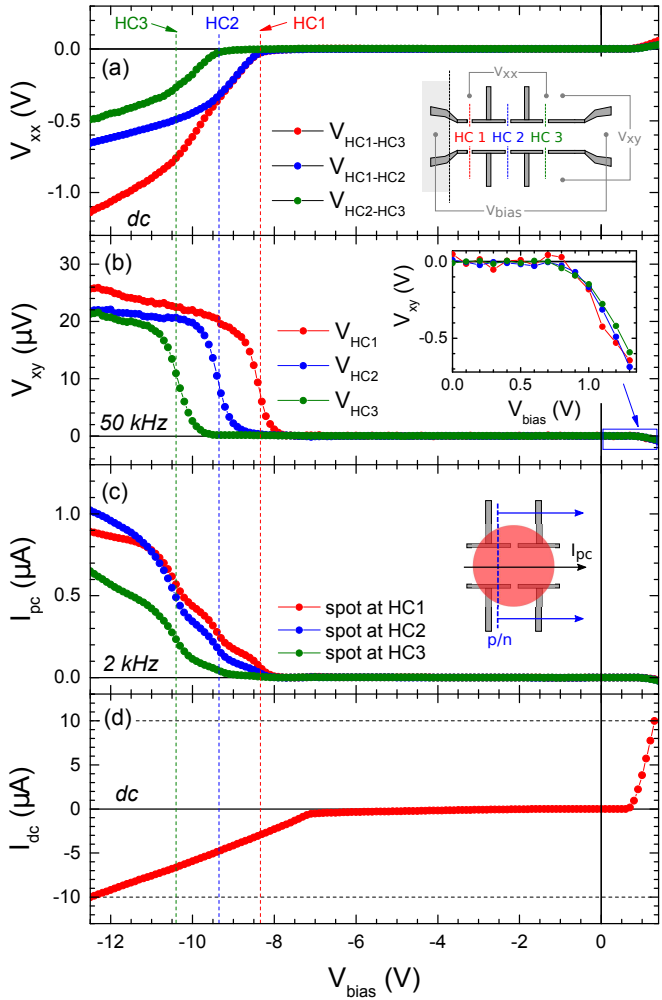}
\caption{\label{fig2} (Color online) (a) The dc longitudinal potential drop $V_{xx}$, measured between the HCs indicated in the panel, with respect to the overall bias $V_{\mathrm{bias}}$. The bias values corresponding to the situation when the pn-front is propagating through each HC are depicted by the vertical dashed lines. The laser spot was positioned on the HC~2. Inset: The scheme of the electrical detection of $V_{xx}$, $V_{xy}$, the dc current $I_{dc}$ and the ac photocurrent $I_{pc}$ (sensed at the reference frequency $f_1$). (b) The ISHE voltages $V_{xy}$ detected at corresponding HCs at the reference frequency $f_2$ with respect to $V_{\mathrm{bias}}$. The laser spot was positioned on the corresponding HC for each measurement.  Inset: The detail of the dependence for $V_{\mathrm{bias}}>0$. (c)~The $I_{pc}$ as a function of $V_{\mathrm{bias}}$ for different laser spot positions. Inset: The sketch explaining the role of the propagation of the depletion front to the overall $I_{pc}$. (d)~The dependence of $I_{dc}$ on $V_{\mathrm{bias}}$, i.e., an I/V characteristic of the pn-junction, with spot located on the HC~2. The horizontal dashed lines represent the maximal current amplitude flowing through the device (10~$\mu$A).}
\end{figure}

It has been shown theoretically and experimentally \cite{gurugubelli2015,reuter2005} that the depletion of planar reverse-bised pn-junctions can exceed 10~$\mu$m, unlike pn-interfaces in bulk systems where the widths of depleted zones is rather in sub-$\mu$m scales. In our device, the range of the expanding depleted region with increasing reverse bias $V_{\mathrm{bias}}$ is detected by sensing the dc longitudinal voltage $V_{xx}$ between HCs along the Hall-bar (Fig.~\ref{fig2}(a) and sketch herein). We observe that for $V_{\mathrm{bias}}>-8V$ the potential drop is located on the pn-junction which does not yet expand towards the HC~1. When $V_{\mathrm{bias}}$ is set below -8~V, however, $V_{xx}$ increases significantly to potential differences of the order of hundreds of mV due to the expansion of the depleted zone through the bar. 
These values represent more than 10$\times$ higher potential drop along the Hall-bar than in the case of $V_{\mathrm{bias}}>0$, if we set the maximal current to $I_{dc}\approx 10$~$\mu$A flowing through the device (compare with Fig.~\ref{fig2}(d)).

The advancing propagation of the depletion zone over the Hall-bar is depicted by $V_{xx}$, measured between different HCs. The potential drop between the HC~1 and HC~3, $V_{\mathrm{ HC1-HC3}}$, shows two changes of its slope for the reverse bias: first, the signal increases rapidly when the edge of the depleted zone expands over the HC~1, and second, when the edge passes over the HC~3 and exits the bar. While the slope after the second change is associated directly with the depleted regime, the slope between the first and the second one is, in addition, affected by the propagation of the depletion edge and represents the transition regime (the observation is also discussed in the following section). Consistently, the potential difference $V_{\mathrm{HC1-HC2}}$ shares the same evolution when the edge passes over the HC~1, but the second change in slope occurs exactly at the HC~2. Analogously, $V_{\mathrm{HC2-HC3}}$ increases when the edge expands into the HC~2 and indicates its exit through the HC~3. The observation allowed us to indicate the position of the depletion edge with respect to a given HC as a function of $V_{\mathrm{bias}}$ (vertical dashed lines in Fig.~\ref{fig2}).

In order to measure the ISHE voltage $V_{xy}$, collected at the three HCs at reference frequency $f_2$ (see Fig.~\ref{fig2}(b) and the inset in Fig.~\ref{fig2}(a)), the circularly polarized light spot is positioned over the corresponding HC which generates locally a spin-polarized current via the optical orientation \cite{agranovich1984}. The ISHE signals $V_{xy}$ are enhanced abruptly when the edge of the depleted zone expands to the corresponding HC, as the spin-polarized photo-current is dramatically increased by the presence of the high potential drop (Fig.~\ref{fig2}(c)). The positions of these steep changes in $V_{xy}$ correspond well with the depletion characteristics seen in Fig.~\ref{fig2}(a).  These successive switchings on the ISHE crosses correspond to a signal amplification by a factor of $\sim30$ at a fixed amplitude of $I_{dc}=10$~$\mu$A (compare the inset in Fig.~\ref{fig2}(b)).

When the depletion-edge passes over a HC, the corresponding $V_{xy}$ tends to rather saturate even if $V_{xx}$ and, thus, the photo-current increase further. This behaviour in similarly high electric fields has been already reported in Ref.~\onlinecite{Miah2007}. Here, the saturation is explained by the reduction of the spin-life time $\tau_s$ due to the enhancement of the Dyakonov-Perel relaxation mechanism for higher $k$-vectors of photo-carriers \cite{kato2004,wu2010}, which plays even a more important role in our higher mobility system.

The motion of the depleted zone edge can be also observed in the comparison of the bias conditions where $V_{xy}$ and $I_{pc}$ start to increase significantly (Figs.~\ref{fig2}(b,c)). The abrupt amplification of $V_{xy}$ tends to occur at higher $V_{\mathrm{bias}}$ then the increase of $I_{pc}$. The effect is well apparent in measurements at the HC~3.
This relative shift with respect to $V_{\mathrm{bias}}$ can be explained if the spin drift length $l_s=E_{xx}\mu\tau_s$ is smaller than the size (FWHM) of the light spot, where $E_{xx}=V_{\mathrm{HC1-HC3}}/6$~$\mu$m is the average electric field in the Hall bar in the depleted regime. Following the inset sketch in Fig.~\ref{fig2}(c), the strong contribution to $I_{pc}$ occurs when the depleted zone edge expands over the illuminated area. The spin-polarized carriers, however, loose their spins before they reach the region the HC is sensitive to (a square of a size of 1~$\mu$m centered at the HC) and do not contribute to the ISHE signal. The $V_{xy}$ is amplified when the depleted zone is expanded further towards the HC, causing a relative shift between $V_{xy}$ and $I_{pc}$. The discussed effect is not strong in measurements at the HC~1, since the HC~1 is located close to the physical edge of the planar pn-junction. In order to get a rough estimate of $\tau_s$, we use the values of $\mu\approx1.6\times10^4$~cm$^2$V$^{-1}$s$^{-1}$, $\mathit{FWHM}\approx 2$~$\mu$m and $E_{xx}\approx10^5$~Vm$^{-1}$, giving $\tau_s<13$~ps. This magnitude of $\tau_s$ supports the assumption of the efficient spin-relaxing mechanism, discussed in the previous paragraph. 

\begin{figure}
\includegraphics[width=1.0\columnwidth]{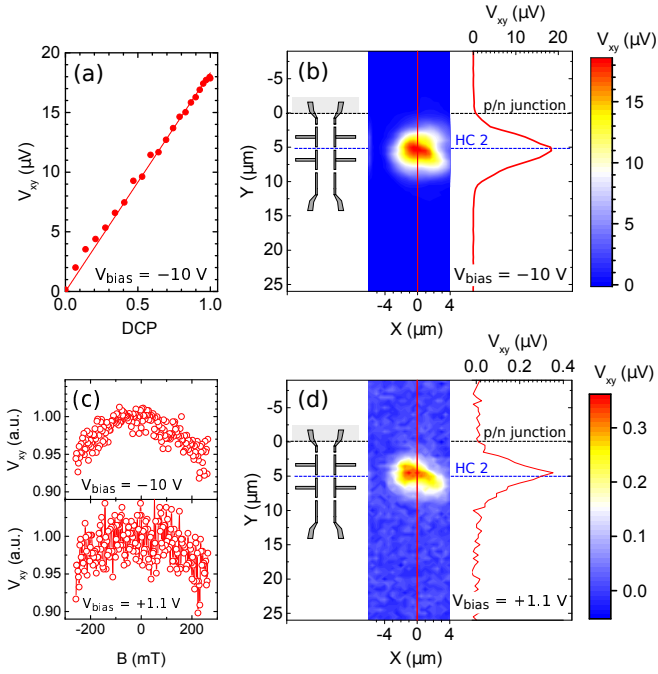}
\caption{\label{fig3} (Color online) (a) The dependence of $V_{xy}$, measured at the HC~2 and $V_{\mathrm{bias}}=-10$~V, with respect to the degree of circular polarization (DPC). The linear trend is a check of the ISHE nature of the measured signals. (b) The 2D response of the HC~2: $V_{xy}$ is measured at HC~2 and $V_{\mathrm{bias}}=-10$~V with respect to the 2D position of the circularly polarized laser spot with the size $\sim2$~$\mu$m. The map shows the local character of the measured signal and no observable non-local (drift/diffusion) contributions. Insets: Sketches of the Hall-bar device (not to scale) and 1D sections along the red line over the 2D maps. (c) Hanle measurements of $V_{xy}$ with respect to the in-plane magnetic field $B$ for $V_{\mathrm{bias}}=-10$~V and $+1.1$~V. (d) The 2D response of the HC~2 to the spot position, measured analogously to (b), for $V_{\mathrm{bias}}=+1.1$~V.}
\end{figure}

More details of the measured ISHE signals are shown in Fig.~\ref{fig3}. The ISHE nature of the signal $V_{xy}$ has been carefully verified by a set of measurements with a varying degree of the circular polarization (DCP) \cite{hecht1987} of the light. The excellent linearity of $V_{xy}$ with DCP, an example of which is shown in Fig.~\ref{fig3}(a) for HC~2 and $V_{\mathrm{bias}}=-10$~V, is a demonstration that the measured signal is free of electrical/optical artifacts.

Considering the above inferred $l_s<\mathit{FWHM}$, the spatial dependence of the ISHE signal is expected to be governed by the Gaussian profile of the laser spot. Indeed, the local character of the ISHE voltage generation can be observed on its 2D spatial dependences (see Fig.~\ref{fig3}(b) and (d)). Here, the $V_{xy}$ was detected at the HC~2 with respect to the 2D position of the laser spot at $V_{\mathrm{bias}}=-10$~V and $+1.1$~V, i.e. $E_{xx}\approx10^5$ and $4\times10^3$~Vm$^{-1}$, respectively. The highly localized, symmetrical and comparable ISHE responses for both bias conditions again illustrate that the spin current is localized within the laser spot-size even at high electric drift fields, confirming that $\tau_s$ is of the order of 10's of ps.
The contribution of a longitudinal diffusive spin transport is not identified, as it would contribute to the 1D profile of the ISHE signal with an odd symmetry with respect to the center of the HC, as shown in Ref.~\onlinecite{nadvornik2015}. 
The estimated magnitude of $\tau_s$ is also consistent with the results of the Hanle measurements at $V_{\mathrm{bias}}=-10$~V and $+1.1$~V (Fig.\ref{fig3}(c)), since the ISHE signal is not significantly reduced in in-plane magnetic fields of the order of tens of mT.

\section{Discussion}
The Joule heating is one of the limiting factors in highly integrated electrical structures. It scales with the dc current as $\sim I_{dc}^2/\sigma$, where $\sigma=en\mu$ is the electrical sheet conductivity and $e$ the elementary electric charge. 
Hence, the efficient suppression of the Joule heating in a uniform channel requires high $\sigma$. 
However, the detected ISHE voltage is \cite{ando2012,bottegoni2013,okamoto2014}
\begin{equation}
\label{eq:SHE}
V_{xy}=\frac{\alpha_{\mathrm{SH}}I_{s}}{\sigma},
\end{equation}
where $\alpha_{\mathrm{SH}}=I_{xy}/I_{s}$ is the spin Hall angle, $I_{xy}$ the transversal charge current due to the ISHE and $I_s$ is the spin-polarized charge current generated by the drift electric field. It means that the high ISHE signal requires low $\sigma$ due to the denominator in Eq.~\ref{eq:SHE} and due to the fact that the low $\sigma$ allows for higher $I_s$. This competing requirements on $\sigma$ can be solved by having a local depletion in the Hall cross area so that the low $\sigma$ is localized at the ISHE detection point while the rest of the transport channel has still high $\sigma$ keeping the total Joule heating low.
We show experimentally this approach using our two-terminal transistor-like device in the results in Fig.~\ref{fig2}(b). Comparing the sensed $V_{xy}$ at a fixed current amplitude $I_{dc}=10$~$\mu$A, the ISHE signal is amplified by more than a factor of 30 in the depleted ($V_{\mathrm{bias}}=-12$~V) regime with respect to the normal regime without the effect of the depletion ($V_{\mathrm{bias}}=+1.3$~V). The latter regime represents the standard detection of $V_{xy}$ in the drift regime, the values of which usually reach $\mu$V or less\cite{Miah2007,nadvornik2015,wunderlich2010}.   

The moderately mobile 2DEG that we use has, compared to low-mobility bulk systems,	 not only a higher $\sigma$  outside the depleted area, but also a smaller $\tau_s$ due to the more effective spin relaxing channels \cite{zhou2007a,wu2010}. $\tau_s$ of the order of tens of ps guarantees, as shown in the 2D spatially resolved responses in Figs.~\ref{fig3}(b) and (d), the local detection of the spin-polarized current even in a drift electric field of the order of $10^5$~Vm$^{-1}$. This combination of locally sensitive and drift-amplified ISHE response in a system with a moderate overall conductivity allows for the application in polarimeter devices with a spatial resolution\cite{nadvornik2015}.

\begin{figure}
\includegraphics[width=1.0\columnwidth]{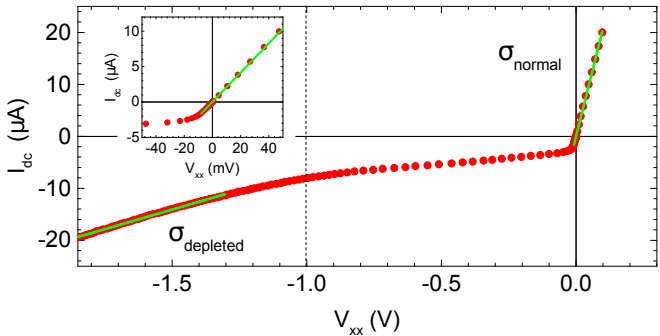}
\caption{\label{fig4} (Color online) The dc current $I_{dc}$ as a function of the dc longitudinal voltage drop $V_{xx}$ between the HC~1 and HC~3. The dependence in the normal and depleted regimes ($V_{xx}>-10$~mV and $V_{xx}<-1.2$~V, respectively) are fitted by linear functions in order to get the corresponding electrical conductivities $\sigma_{\mathrm{normal}}$ and $\sigma_{\mathrm{depleted}}$ (fits are depicted by the green lines). Inset: A detail for $V_{xx}>-50$~mV.}
\end{figure}

The change of the carrier concentration, the key functionality of the device, should also affect the spin-transport characteristics. Namely, $\alpha_{\mathrm{SH}}$ is expected to decrease with decreasing carrier concentration according to Ref.~\onlinecite{matsuzaka2009}. To demonstrate this, we need to infer $\alpha_{\mathrm{SH}}$ from Eq.~\ref{eq:SHE} in the normal and depleted regime. The corresponding $\sigma$ in both regimes is inferred from Fig.~\ref{fig4} which displays the evolution of $I_{dc}$ as a function of $V_{xx}$, i.e., the I/V characteristic of the Hall bar. It clearly shows the high conductive normal regime, unaffected by the pn-junction, with $\sigma_{\mathrm{normal}}=1.2\times10^{-3}$~$\Omega^{-1}$. When the depleted zone edge is expanding through the Hall bar, the conductivity gradually reduces. After the edge exits completely the Hall bar at $V_{\mathrm{bias}} < -1.2$~V the conductivity saturates at $\sigma_{\mathrm{depleted}}=9.0\times10^{-5}$~$\Omega^{-1}$ at $V_{\mathrm{bias}}<-1.2$~V. We use these two values of $\sigma$ to describe the corresponding regimes.


In order to evaluate $\alpha_{\mathrm{SH}}$ from Eq.~\ref{eq:SHE} we have to determine $I_s$ in terms of the photo-generated current $I_{pc}$. We define $I_s=en_sv$ as the flux of the spin-polarized carriers with drift velocity $v$ and with density $n_s=n^{\uparrow}-n^{\downarrow}$, where $n^{\uparrow}$ and $n^{\downarrow}$ are the total carrier densities of corresponding spins. We can, thus, express 
\begin{equation}
I_s=I_{pc}\frac{n_s}{n_p},
\label{eq:Is}
\end{equation}
where $n_p$ is the concentration of the photo-generated carriers. The steady state $n_s$ and $n_p$ upon a continuous illumination with the photo-carrier generation rate $G$ follow from the steady state solutions of the rate equations \cite{okamoto2014,agranovich1984,dyakonov2008}, i. e., $n_s=P_0G\tau_s$ and $n_p=G\tau$. Here, $P_0$ is the degree of the spin-polarization of the photo-generated electrons and $\tau$ is the photo-carrier recombination time. In case of a doped semiconductor $\tau_s$ is not limited by  $\tau$ since the recombination process involves also the unpolarized electrons from the equilibrium electron concentration in dark \cite{dzhioev2002b,kikkawa1998,salis2010}. This is why $n_s$ can reach higher values than $n_p$ in doped systems.


As the $\tau$ is not experimentally known, we first evaluate the effective spin Hall angle $\tilde{\alpha}_{\mathrm{SH}}$, defined from Eqs.~\ref{eq:SHE} and \ref{eq:Is} as
\begin{equation}
\tilde{\alpha}_{\mathrm{SH}}=\alpha_{\mathrm{SH}}\frac{\tau_s}{\tau}=\frac{V_{xy}\sigma}{I_{pc}P_0}.
\label{eq:SHA}
\end{equation}
We use $V_{xy}$, $I_{pc}$ and $\sigma=\sigma_{\mathrm{depleted}}$ and $\sigma_{\mathrm{normal}}$ corresponding to the fully depleted and normal regimes at $V_{\mathrm{bias}}=-12$~V and $V_{\mathrm{bias}}=+1.3$~V, respectively. We assume $P_0=1$ for both the depleted and undepleted regimes, which is the maximum degree of the optically injected spin-polarization in 2DEGs at the instant of the photo-generation\cite{dareys1993,pfalz2005,agranovich1984}. We then get the lower bound of $\tilde{\alpha}_{\mathrm{SH}}^{\mathrm{depleted}} = (2.7\pm 0.6)\times10^{-3}$ and $\tilde{\alpha}_{\mathrm{SH}}^{\mathrm{normal}} = (7\pm 2)\times10^{-2}$. The observed decrease of the effective SHA by one order of magnitude in the depleted regime is consistent with the expected behaviour from Ref.~\onlinecite{matsuzaka2009}, where the same suppression of $\alpha_{\mathrm{SH}}$ is reported over one order of magnitude change in concentration. Considering $\sigma_{\mathrm{normal}}/ \sigma_{\mathrm{depleted}}\approx10$, the carrier concentration in our case also changes by an order of magnitude. In addition, the relative reduction of $\tilde{\alpha}_{\mathrm{SH}}^{\mathrm{depleted}}$ with respect to the normal regime competes with the drift effect on the ISHE signal upon the depletion, suggesting that further optimization of transport parameters could bring even a more efficient amplification of the electrical ISHE signal in the depleted regime.

Values of $\alpha_{\mathrm{SH}}$ reported in bulk and similar low dimensional systems \cite{wunderlich2009,olejnik2012,matsuzaka2009,garlid2010} do not usually exceed $10^{-2}$. This is consistent with our values of $\tilde{\alpha}_{\mathrm{SH}}$ if $\tau<\tau_s\sim10$~ps, which would be an indication of highly effective photo-carrier recombination. These $\tau$ values can be found in GaAs at low temperatures, especially if trapping processes are significant \cite{fukumoto2015a,gupta1991,mcintosh1997,prabhu1997,smith1997}.

\section{Conclusion}
In conclusion, we have demonstrated that the large, bias-controlled expansion of the depleted zone in a lateral pn-junction can be used to amplify the electrically detected ISHE signals by more than one order of magnitude with respect to the unbiased regime. In this two-terminal transistor-like device, the source and drain contacts are used both to apply the drift bias and to deplete the carrier density in the Hall bar by the expanding depletion zone, whose width is spatially controlled with sub-$\mu$m resolution. Due to the low-dimensional nature of the sample, the reduced spin life-time and the corresponding spin drift-length allow us to perform local detection of spin currents even in high electric field of the order of $10^5$~Vm$^{-1}$. The gating-like effect of the device affects the spin Hall angle consistently with the previous reports and gives the perspective of a further optimization of the spin Hall device. The combination of the ISHE signal amplification by the drift and depletion functionalities, together with the local character of the spin detection guaranteed in wide range of drift conditions, is promising for concepts of ISHE-based spintronic devices, such as spintronic polarimeters.


\begin{acknowledgments}
We acknowledge support from the European Research Council (ERC) Advanced Grant No. 268066, from
European Metrology Research Programme within the Joint Research Project EXL04 (SpinCal), from the Ministry
of Education of the Czech Republic Grant No. LM2015087, from the Czech Science Foundation Grant No. 14-37427G, from the Charles University Grants No. 1360313 and No. SVV-2015-260216.
\end{acknowledgments}
 

%
%


\providecommand{\noopsort}[1]{}\providecommand{\singleletter}[1]{#1}%

\end{document}